\documentclass[twocolumn,superscriptaddress,longbibliography,aps,pra,floatfix]{revtex4-2}

\usepackage{graphicx}
\usepackage{bm}
\usepackage{color}
\usepackage{amsmath}
\usepackage{amssymb}
\usepackage{epstopdf}
\usepackage{appendix}

\usepackage{enumitem}

\usepackage{float}

\usepackage{xcolor}

\usepackage{ulem}

\usepackage[urlcolor=blue,colorlinks=true,citecolor=blue,linkcolor=blue,pdfstartview={FitH},bookmarks=false]{hyperref}

\definecolor{mygreen}{RGB}{0, 102, 0}

\begin{document}

\title{ Density functional theory study of Au-\textit{fcc}/Ge and Au-\textit{hcp}/Ge interfaces}

\author{Olga~Sikora}
\email{olga.sikora@pk.edu.pl}
\affiliation{\mbox{Faculty of Materials Engineering and Physics, Cracow University of Technology, Podchor\c{a}\.zych 1, PL-30084 Krak\'ow, Poland}}

\author{Ma\l{}gorzata~Sternik}
\affiliation{\mbox{Institute of Nuclear Physics, Polish Academy of Sciences,
W. E. Radzikowskiego 152, PL-31342 Krak\'{o}w, Poland}}

\author{Benedykt R. Jany}
\affiliation{
Marian Smoluchowski Institute of Physics, Faculty of Physics, Astronomy and Applied Computer Science, Jagiellonian University, {\L}ojasiewicza 11, 30348 Krakow, Poland}

\author{Franciszek Krok}
\affiliation{
Marian Smoluchowski Institute of Physics, Faculty of Physics, Astronomy and Applied Computer Science, Jagiellonian University, {\L}ojasiewicza 11, 30348 Krakow, Poland}

\author{Przemys\l{}aw~Piekarz}
\affiliation{\mbox{Institute of Nuclear Physics, Polish Academy of Sciences,
W. E. Radzikowskiego 152, PL-31342 Krak\'{o}w, Poland}}

\author{Andrzej~M.~Ole\'s$\,$}
\affiliation{Max Planck Institute for Solid State Research, Heisenbergstrasse 1,
D-70569 Stuttgart, Germany}
\affiliation{\mbox{Institute of Theoretical Physics, Jagiellonian University,
Prof. Stanis\l{}awa {\L}ojasiewicza 11, PL-30348 Krak\'ow, Poland}}

\date{\today}

\begin{abstract}
In recent years, nanostructures with hexagonal polytypes of gold have been synthesised, opening new possibilities in nanoscience and technology.
As bulk gold crystallizes in the \textit{fcc} phase, surface effects can play an important role in stabilizing hexagonal gold nanostructures.
Here we investigate several hetero-structures with Ge substrate, including  the \textit{fcc} and \textit{hcp} phases of gold that have been observed experimentally.
We determine and discuss their interfacial energies and optimized atomic arrangements, comparing the theory results with available experimental data.
Our DFT calculations for the Au-\textit{fcc}(011)/Ge(001) junction show how the presence of defects in the interface layer can help to stabilize the atomic pattern consistent with microscopic images.
Although the Au-\textit{hcp}/Ge interface is characterized by similar interface energy, it reveals large atomic displacements due to significant mismatch.
Finally, analyzing the electronic properties, we demonstrate that
Au/Ge systems have metallic character but covalent-like bonding states between interfacial Ge and Au atoms are also present.
\end{abstract}

\maketitle

\section{Introduction}
\label{sec.intro}
Heterophase interfaces are responsible for unique properties of many advanced devices designed for electronics and other applications \cite{Butler_2019}.
Understanding the formation and energetics of interfaces is highly important for the nucleation of new crystalline phases on a specific substrate \cite{Goune_2015, Shi_2012}.
The structure of a heterophase can be studied using advanced atomic-resolution experiments, such as high-resolution electron microscopy \cite{merkle1992}, high-resolution secondary-electron microscopy \cite{ciston2015}, scanning transmission electron microscopy  \cite{nellist2004, guzzinati2018}, or scanning-tunnelling microscopy.
However, experimental data provide only partial insight into the formation of interfaces and the interpretation of microscopic images is often ambiguous. Various simulation techniques can be used to resolve the uncertainties and reveal mechanisms stabilizing the observed interfaces.
Deeper knowledge  of surfaces and interfaces in hetero-structures can play a crucial role in developing new methods for synthesizing such materials and in expanding their possible applications in nanoscience and technology.
As a result of advances in computational methods and the increasing computer power, the atomic structure and specific properties of various interfaces have been successfully studied using first principle calculations \mbox{\cite{Benedek2000, Benedek2002, Wang2005, Matsunaka2008, Lu2013, Li2022}.}

Up to very high pressure, gold crystallizes in the \textit{fcc} phase ~\cite{2007_dubrovinsky} with the $ABC$ repeating pattern of hexagonal planes in the (cubic) [111] direction.
Recently, nanostructures including different polytypes of gold have been observed in various experiments and have gained interest due to their potential applications. For example, the interface regions of polytypic gold nanorods were found to be highly active in catalysis~\cite{2020_fan}.
 The hexagonal phases seen in experiments are characterized by $ABAB$ (\textit{hcp}) or $ABAC$ (\textit{dhcp}) stacking patterns.
The \textit{hcp} surface has been observed experimentally in nanowires~\cite{1997_kondo} and ultrathin sheets on graphene oxide~\cite{2011_huang}.
Nanoribbons with metastable \textit{dhcp} structure have been also reported~\cite{2014_fan, 2017_fan} and were used to grow the \textit{dhcp} forms of several other metals. It is worth mentioning that also stable hexagonal silver (\textit{dhcp}) nanostructures have been syntesized~\cite{2001_taneja, 2006_liu}.

Nanostructures of \textit{hcp} gold were found after growing Ge nanowires with Au as catalyst~\cite{2010_marschall}, and a  possible mechanism responsible for the formation of \textit{hcp} gold has been suggested, involving hexagonal AuGe $\beta$-phase present at the intermediate stages of growth.
Interestingly, while the \textit{fcc} crystallites were randomly oriented with respect to the Ge substrate, the \textit{hcp} nanostructures were typically found with (001) planes at 60-65$^{\circ}$ to the (111) Ge planes~\cite{2010_marschall}.
In a recent experiment, stable \textit{hcp} nanoislands we obtained under controlled annealing conditions on the germanium substrate~\cite{2017_jany}. After initial crystallization of the \textit{fcc} gold phase, the \textit{hcp} phase grows from the eutectic Au/Ge liquid. Again, there seems to be a preferred \textit{hcp} crystal orientation with the Au(010) plane, or Au(01$\bar{1}$0) in the Miller-Bravais notation, parallel to the Ge(111) plane.

First principles calculations of the cohesive energy and elastic constants as well as phonon dispersion relations show stability of both \textit{hcp} and \textit{dhcp} polytypes of gold~\cite{2017_benaissa}. \textit{Ab initio} studies indicate higher stability of the \textit{fcc} phase and a tendency towards \textit{hcp} $\to$  \textit{\textit{fcc}} phase transformation,
however, the calculated differences between the polytypes are very small~\cite{2007_dubrovinsky, 2015_wang}. On the other hand, molecular dynamics simulations of metallic nanowires show a phase transformation from \textit{fcc} to \textit{hcp} below a critical diameter~\cite{2011_sutrakar}.
These findings support feasibility of obtaining new hexagonal structures under conditions where surface effects are significant.
The purpose of this paper is to 
provide a quantitative decription of the energetics of interface and to draw possible conclusions concerning the type and formation of Au/Ge hetero-structures. Therefore, the concept of interfacial energy and of the work of separation (adhesion) were used.
We applied first-principles methods based on density functional theory (DFT) to study Au/Ge hetero-structures with different interfacial plane orientations.

The remaining of the paper is organized as follows.
The methodology and the calculation details are given in Sec.~\ref{sec.method}~A, and in Sec.~\ref{sec.method}~B the investigated crystal planes and possible interfaces are introduced.
Convergence tests presented in Sec.~\ref{sec.method}~C are conducted for the simplest, almost strain-free Au-\textit{fcc}(001)$\parallel$Ge(001) hetero-structure.
Section~\ref{sec.method}~D begins from discussing the calculated surface energies for several Au and Ge planes as well as  available experimental data. In the rest of this Section two simple Au/Ge interfaces are discussed.
In Sec.~\ref{sec.result}~A we compare several variants of a hetero-structure with parallel Ge(001) and Au-\textit{fcc}(011) planes and their different mutual position. To the best of our knowledge, only one of the structures, with
[110]Ge$\parallel[0\bar{1}1]$Au-\textit{fcc}, has been observed~\cite{2017_jany}.
We show the optimal configuration obtained from the calculation and discuss the experimental findings.
In Sec.~\ref{sec.result}~B we investigate the novel $hcp$ structure that is adjacent to the Ge(111) surface. From the electron microscope image~\cite{2017_jany} one can identify the Au-$hcp$ plane parallel to Ge substrate as $(010)$. We discuss the optimized structures and consider defects that could stabilize the interface.
In Sec.~\ref{sec.electronic} we analyse the electronic properties of Au/Ge systems and the formation of Ge-Au bonds.
The paper is concluded in Sec.~\ref{sec.summa}.
The Appendices provide surface energy details for gold and germanium crystals (\ref{sec.appA}) and the charge density differences at the Au/Ge
interface (\ref{sec.appB}).

\section{Methodology}
\label{sec.method}
\subsection{Interfacial energy and the work of separation\\ in DFT calculations}

In contrast to the bulk phase of a material, the surface atoms have an incomplete set of neighbors and therefore unrealized bonding energy (surface energy).
When two crystalline solids bind together, the atoms or molecules form an interface with preferred plane orientation relationship determined by
the strength of bonding, which can be quantified using the concepts of interfacial free energy and the work of separation.
Their values can be obtained from the DFT calculations of total energies of the appropriate systems modeling the hetero-structure as well as the bulk crystals and slabs with a vacuum layer.
The methodology details (from choosing the interface model up to the optimization methods) differ between implementations \cite{Lu2013, Xu2015}.
Here we focus on the approach in which two crystalline solids ($X$ and $Y$) form a supercell with imposed periodic boundary conditions, i.e. there are two interfaces within the supercell. The interfacial energy $\gamma_{int}$ is defined as:
\begin{equation}
\gamma_{int} = \frac{1}{2A_{int}} \left[E_{XY} - \left(n_{X}\varepsilon_{X}^{bulk} + n_{Y}\varepsilon_{Y}^{bulk}\right)\right],
\end{equation}
where $E_{XY}$ is the total energy of the \textit{XY} hetero-structure, $\varepsilon_{X(Y)}^{bulk}$ is the energy per atom in the $X(Y)$ bulk form, $n_{X(Y)}$ is the number of $X(Y)$ atoms in the \textit{XY} supercell, and $A_{int}$ is the area of the interface.

The work of separation ($W_{sep}$) is defined as the energy required to reversibly separate a bulk material into two semi-infinite bulks with free surfaces. Here we calculate this energy for an $XY$ hetero-structure, separated into two slabs, i.e., two supercells with either $X$ or $Y$ atoms and a vacuum layer, with periodic boundary conditions imposed:
\begin{equation}
W_{sep} =   \sigma_X + \sigma_Y - \gamma_{int},
\end{equation}
where $\sigma_X$ and $\sigma_Y$ are the surface energies of $X$ and $Y$ phases. Their values can be obtained from the following formula:
\begin{equation}
\sigma_{X(Y)} = \frac{1}{2A_{int}} \left(E_{X(Y)}^{slab} - n_{X(Y)} \varepsilon_{X(Y)}^{bulk}\right), \end{equation}
where $E_X(Y)^{slab}$ is the total energy of the $X(Y)$ slab with two free surfaces. %

All calculations were based on the DFT method implemented in the plane wave basis VASP code \cite{VASP1, VASP2}.
The Perdew-Burke-Ernzerhof (PBE) functional \cite{PBE} within the generalized gradient approximation (GGA) for the exchange and correlation energy was used.
The electronic wave functions were expanded as linear combinations of plane waves, truncated to include only plane waves with kinetic energies below a cutoff energy $E_\text{cut} = 350$~eV.
The valence states were optimized with the Ge($s^{2}p^{2}$) and Au($s^1d^{10}$) electron configurations.

Bulk materials, surfaces and interfaces were simulated using the supercell method with the periodic boundary conditions.
Optimizations of the structural parameters (lattice constants and atomic positions) were carried out using the Monkhorst-Pack grid of {\bf k}-points appropriate for the calculated structure: from dense (8,8,8) grid for crystallographic cells of bulk crystal to (4,4,4) grid for larger superlattices that model the Au-\textit{fcc}(011)/Ge(001) and Au-\textit{hcp}(010)/Ge(111)  hetero-structures.
The slabs with vacuum are calculated using (8,8,2) or (4,4,2) {\bf k}-points grids depending on the size of the cross-section.
A vacuum layer of 15~\AA\ was selected for each structure to eliminate interactions between the two surfaces.
The conditions for ending the optimization loops for electronic and ionic degrees of freedom were defined by the total energy difference between steps of 10$^{-8}$ eV and the internal forces of 10$^{-2}$ eV\AA$^{-1}$, respectively.

\begin{figure}[t!]
\centering
\includegraphics[width=\linewidth]{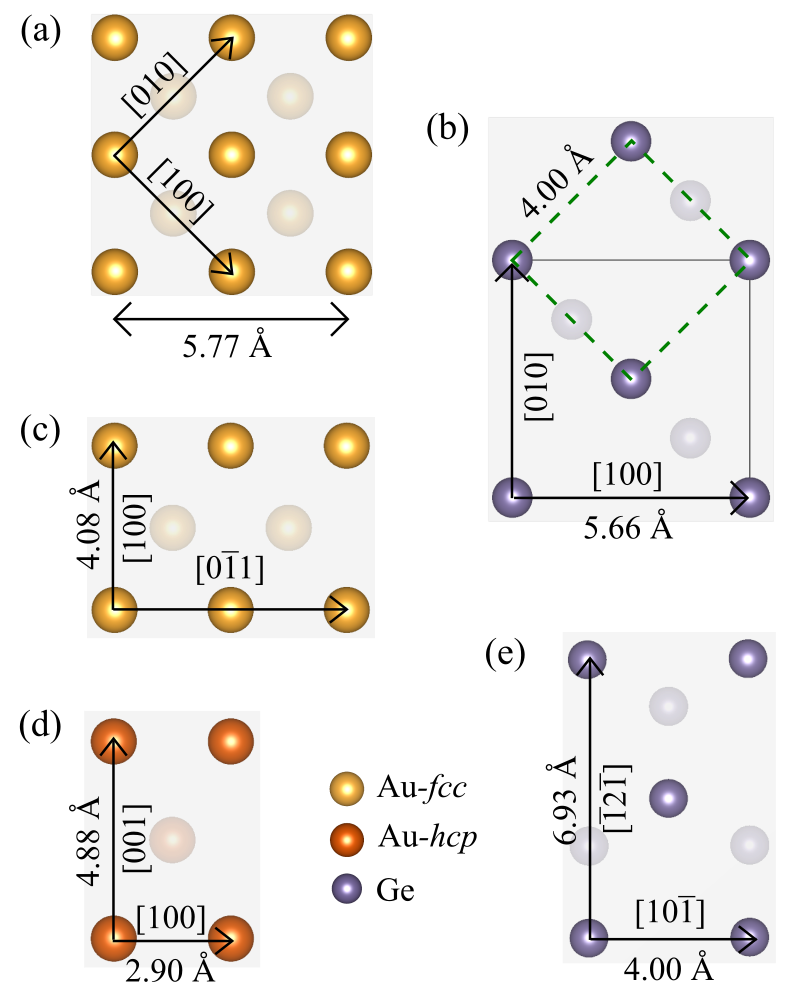}
\caption{Surfaces of Au and Ge slabs investigated in our study, with experimental interatomic distances: \mbox{(a) Au-{\textit fcc}(001);} (b) Ge(001), dashed green square indicates the building blocks matching the [100]Au length; (c) Au-{\textit fcc}(011); \mbox{(d) Au-{\textit hcp}$(010)$,} or $(01\bar10)$ in the Miller-Bravais notation; \mbox{(e) Ge(111)}. To distinguish the \textit{hcp} lattice, the Au atoms are shown in dark orange.
}
\label{fig.basic}
\end{figure}

In all calculations, the starting configurations (Au/Ge supercells, bulk structures and isolated slabs) were prepared under assumption that the lateral extent of the golden part of heterostructures match the optimized dimensions of the germanium lattice, treated here as a substrate.
The calculated interfacial energies capture therefore mainly the chemical bonding energy at the interface and the energy cost of internal distortions, as the strain coming from the lattice mismatch is present all investigated configurations.
The lattice constant perpendicular to the interface as well as the atomic positions are fully relaxed.

\subsection{Structural models}

Under ambient pressure, germanium and gold crystallize in the diamond ($Fd\bar{3}m$) and the $fcc$ ($Fm3m$) structure, respectively.
Their experimental lattice constants are $a_{\textnormal{Ge}}=5.66$~\AA\ for germanium and $a_{\textnormal{Au-}fcc}=4.08$~\AA\ for gold \cite{kittel_lattices}.
The hexagonal (\textit{hcp}) phase of Au formed during the specific process of annealing and cooling of gold and germanium is characterized by the $P6_3/mmc$ space group with lattice constants $a_{\textnormal{Au-}hcp}=2.90$~\AA~and $c_{\textnormal{Au-}hcp}=4.88$~\AA~\cite{2017_jany}.
These structural parameters are well reproduced in our DFT calculations.
Obtained cubic lattice constants are equal to 5.77~\AA~ and 4.16~\AA~ for the Ge and Au-\textit{fcc}, respectively, and the hexagonal lattice parameters are 2.93 \AA~ and 4.89 \AA~ for Au-\textit{hcp}.

Figure~\ref{fig.basic} shows surfaces that terminate slabs which are the building blocks for hetero-structures investigated in this study.
We constructed them by looking for matching lattice parameters and following the experimental data if available. Moreover, different mutual positions of the slabs were considered in order to find the best atomic arrangement.

It can be easily noticed that (001) faces of Au-\textit{fcc} and Ge can form almost strain-free hetero-structure if the gold slab as in panel (a) is joined with the Ge slab marked with a black square in panel (b).
A variant of such superlattice was used to perform convergence tests described in the next subsection.

Another way of exploiting this lattice match is to connect Au-\textit{fcc}(011) plane as shown in panel (c) and the same Ge(001) slab. We note that now we get a strain-free interface only in one direction, and we need a few Au and Ge slabs to reduce the lattice strain in the the perpendicular direction. Alternatively, the  Au-\textit{fcc}(011)/Ge(001) hetero-structure can be built using a Ge slab marked with the green square, as its side length matches the gold lattice constant. Again, multiple cells are needed in the perpendicular direction.

There is no obvious way to build the Au-\textit{hcp}/Ge interface using the low-index Au-\textit{hcp} and Ge surfaces, so we decided to make calculations for the experimentally found Au-\textit{hcp}$(010)$/Ge(111) interface only. To model the hetero-structure, we multiply  the relevant Au and Ge surfaces [shown in panels (d-e)] to get the minimal strains within a reasonable size of the system.

\subsection{Convergence tests}

\begin{figure}[t!]
\centering
\includegraphics[width=\linewidth]{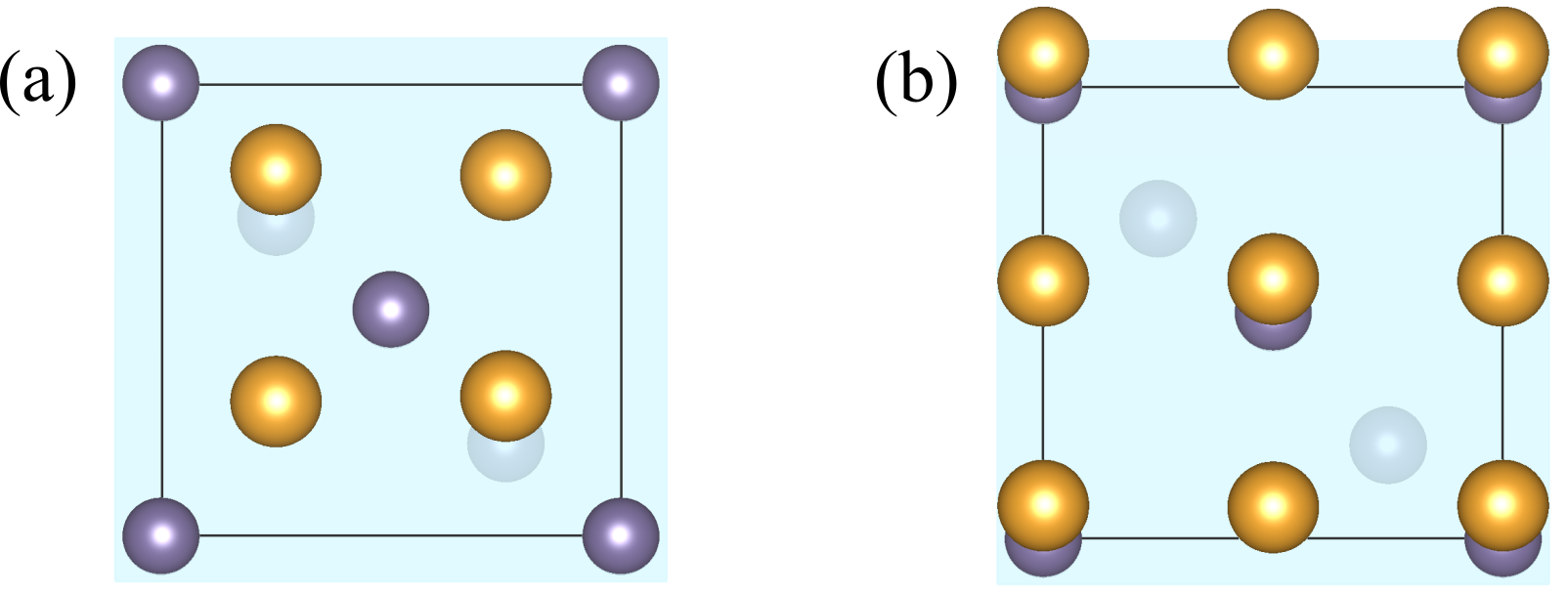}
\caption{Two variants of Au-\textit{fcc}(001)/Ge(001) interface atomic arrangement: (a) variant $T_1$ with the interfacial Au atoms above bridge sites of Ge surface~--~this structure was used for convergence tests, and (b) variant $T_2$ with Au atoms directly above the surface Ge atoms. The shaded surface indicates the interface Ge layer, and the second Ge layer is visible below (crystal planes are slightly tilted to show all the atoms).
}
\label{fig.ini_slabs}
\end{figure}

Ideally, the thickness of the slabs should be large enough to ensure that all the surface effects are captured within the supercell and that the two interfaces resulting from periodic boundary conditions do not interact with each other.
To test the convergence of surface energies, interface energies and the work of separation we consi\-dered the Ge(001) and Au-\textit{fcc}(001) slabs with different numbers of atomic layers and the epitaxial Au-\textit{fcc}(001)/Ge(001) heterojunction shown in Fig.~\ref{fig.ini_slabs} (two variants of mutual positions of the slabs are presented).
This simplest hetero-structure can be built by setting the [110] direction of the Au crystal parallel to the [100] direction of the Ge lattice~\cite{Gerbi2018}, i.e. combining two fragments of the lattice planes oriented as in Fig.~\ref{fig.basic}~(a) and (b).
The resulting mismatch $\epsilon$ defined as:
\begin{equation}
\epsilon =\left(a_\text{Au-fcc}\sqrt2 - a_\text{Ge}\right)/a_\text{Ge}\times 100\%,
\end{equation}
is equal to only 2.0\%, i.e., the hetero-structure is almost strain-free.
As only four Au atoms and two Ge atoms form a single layer of this interface, one can easily increase the number of layers in the supercell.
In variant $T_1$ of the considered hetero-structure, the Au atoms are maximally away from the Ge atoms, while in variant $T_2$ some of the gold atoms are located on the top of Ge sites.
\begin{table*}[t!]
\caption{Interfacial energies ($\gamma_{int}$), work of separation values ($W_{sep}$), rumpling parameters $r_\text{Ge}$ ($r_\text{Au}$) for interfacial Ge (Au) layers, and interlayer distances at the interface ($d_{int}$) calculated for different Au/Ge hetero-structures as well as Au-fcc/Au-\textit{hcp} superlattice.}
\label{tab.001}
\begin{ruledtabular}
\begin{tabular}{l c c c c c c c}
interface&variant~~~&~~~$\gamma_{int}$ (J/m$^2$)~~~~~&~~~$W_{sep}$(J/m$^2$)~~~~~& $~~~r_\text{Ge}$ (\AA)~~~&~~ $~~~r_\text{Au}$ (\AA)~~~~~& $~~~d_{int}$ (\AA)~~~~~\\
\hline
Au-\textit{fcc}(001)/Ge(001) & $T_1$ & 0.345 & 1.862  & 0.00 & 0.40  & 1.76 \\
                             & $T_2$ & 0.909 & 1.298  & 0.00 & 0.13  & 2.36 \\
\hline
Au-\textit{fcc}(111)/Au-\textit{hcp}(001)~~~ & -- & 0.041 & 1.411 & -- & 0.00 & 2.41 \\
\hline
Au-\textit{fcc}(011)/Ge(001) & $A$ & 0.590 & 1.638 & 0.08 & 0.53 & 1.55 \\
                             & $B$ & 0.821 & 1.407 & 0.49 & 0.38 & 1.93 \\
                             & $C$ & 0.336 & 1.892 & 0.21 & 0.61 & 1.77 \\
                             & $D$ & 0.437 & 1.784 & 0.29 & 0.62 & 2.13 \\
\hline
Au-\textit{hcp}(010)/Ge(111) & -- & 0.369  & 1.506 & 0.23 & 0.30 & 2.17 \\
\end{tabular}
\end{ruledtabular}

\end{table*}

\begin{figure}[t!]
\centering
\includegraphics[width=\linewidth]{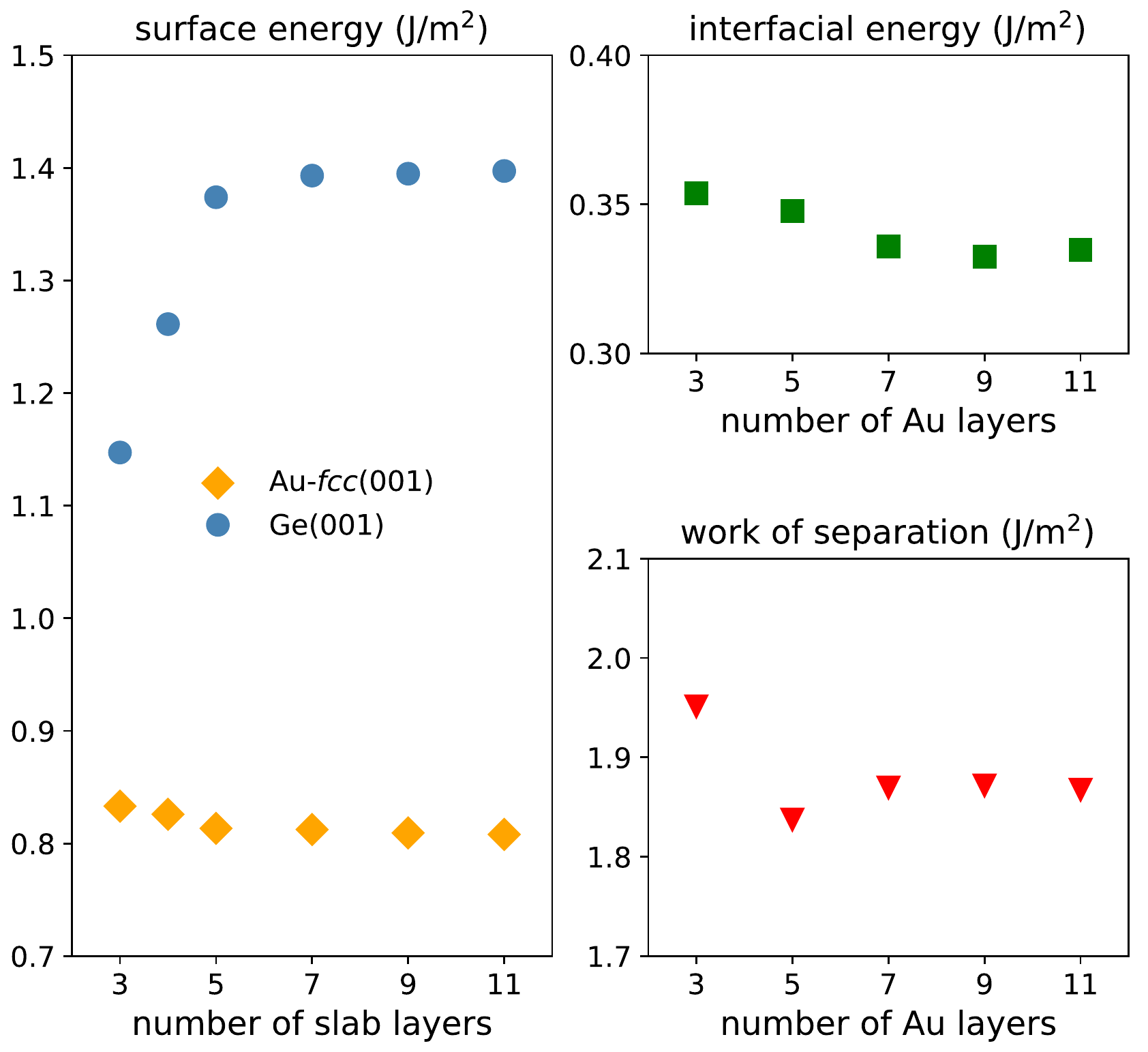}
\caption{
Convergence tests: (left) dependence of Au-\textit{fcc}(001) and Ge(001) surface energies on the number of slab layers, (right) the interfacial energy and work of separation for Au\text{fcc}(001)/Ge(001) hetero-junction vs. the number of Au layers (the number of Ge layers is equal to five).
}
\label{fig.conv}
\end{figure}

Our preliminary calculations indicated that variant $T_1$ is preferred and for this structure we conducted convergence tests.
First, separated slabs of Ge and Au-\textit{fcc} crystals terminated with (001) plane and vacuum layer of 15~\AA~  were used to determine the optimal slab thicknesses for surface energy calculations.
The calculated values plotted in
Fig.~\ref{fig.conv}~(left panel) show that above the fifth layer the discrepancies between results are smaller than 0.03~J/m$^2$. Large initial increase of the surface energy values seen in the data for Ge crystal results from very small number of atoms (only six) in this slab, and two free surfaces.

Right panels of Fig.~\ref{fig.conv} present the convergence tests performed to find the optimal thickness of the Au layer in the Au/Ge hetero-structure with five Ge layers.
The plot of the interfacial energy reveals much weaker dependence on the number of Au layers.
Values of the work of separation for Au-\textit{fcc}(001)/Ge(001) hetero-junction include both the calculated surface and interface energies, therefore again we observe differences in the results of about 0.03 J/m$^2$ for slabs with more than five layers of Au.
We conclude that five layers of both Au and Ge atoms in the slabs are sufficient for our calculations. This observation is important for the investigation of more complicated interfaces where larger numbers of layers result in a very long computational time.

\subsection{Results for model surfaces and interfaces}

\begin{figure}[b!]
\centering
\includegraphics[width=\linewidth]{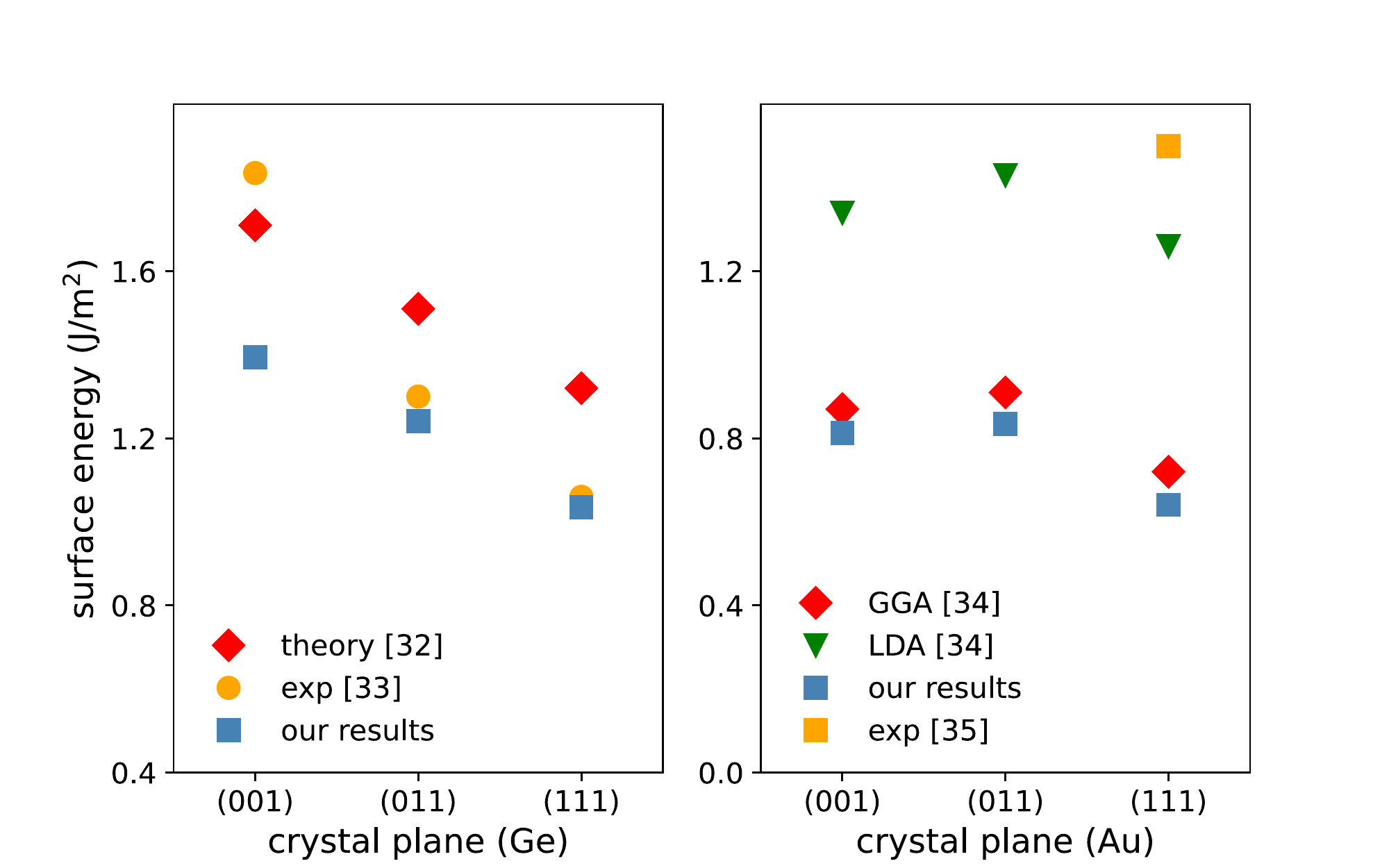}
\caption{
Surface energies calculated for germanium (left) and \textit{fcc}-gold (right) crystal planes compared to available experimental and theory data (values are given in Appendix~\ref{sec.appA}).
}
\label{fig.surf_en}
\end{figure}
We consider several low-index planes of Ge, Au-\textit{fcc} and Au-\textit{hcp} crystals
(some of them are presented in Fig.~\ref{fig.basic}).
The calculated surface energies are presented in (Fig.~\ref{fig.surf_en}) and Appendix~\ref{sec.appA}.
The comparison of our results with the available theoretical and experimental data shows that different theoretical approaches lead to discrepancies in absolute values of surface energies, however, the relative changes between the investigated crystal planes are very similar.
Experimental data for Ge crystal show the same trend, i.e. lowering the surface energy from (001) to (111) plane, with a slightly bigger slope.

\begin{figure}[b!]
\centering
\includegraphics[width=1.07\linewidth]{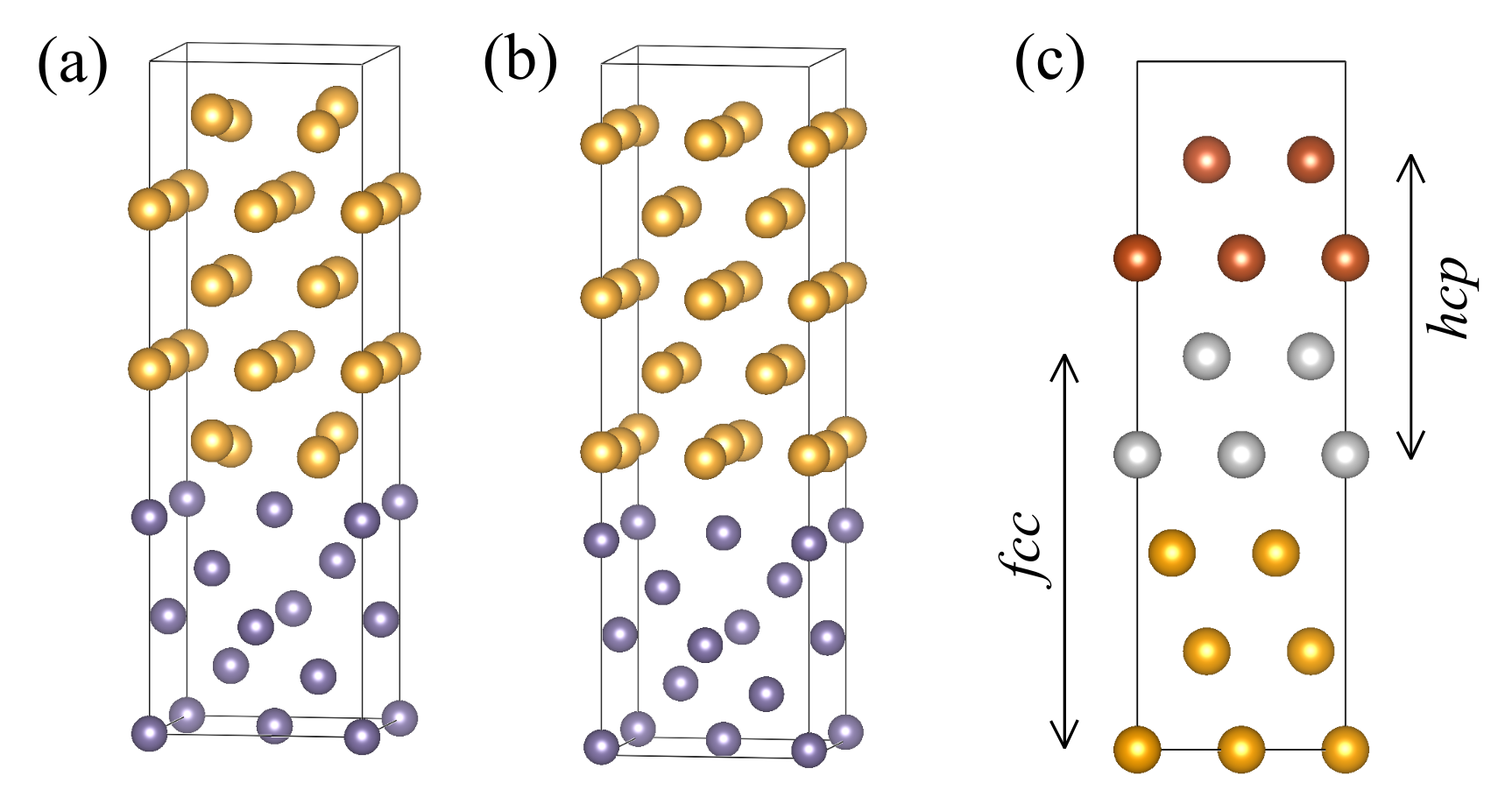}
\caption{
The relaxed structures of the simplest coherent interfaces. Two variants of the Au-\textit{fcc}(001)/Ge(001) hetero-structure: (a) variant $T_1$, with interfacial Au atoms at bridge sites of Ge surface, and (b)  variant $T_2$, with interfacial Au atoms at top sites of Ge surface.
Panel (c) shows the Au-\textit{fcc}(001)/Au-\textit{hcp}(0001) interface with the two phases of gold marked in yellow and dark orange, respectively. The positions of the gray atoms are compatible with both arrangements.
}
\label{fig.s001}
\end{figure}

The optimized atomic arrangements for our model Au-\textit{fcc}(001)/Ge(001) hetero-structures are shown in Fig.~\ref{fig.s001} (variant $T_1$ and $T_2$).
The total energies of two considered supercells are $-108.806$~eV and $-106.484$~eV for the variant $T_1$ and $T_2$, respectively. Therefore, the realization of the variant $T_1$ with interfacial Au atoms located at bridge sites of germanium surface is more probable.
In Table~\ref{tab.001} we present the calculated interface energy and work of separation together with the structural parameters useful for describing interfaces: (i) the intralayer rumpling parameter ($r_{Ge}$ and $r_{Au}$), defined as the maximal difference between the $z^{th}$ coordinates of atoms belonging to one layer, (ii)~the interlayer distance ($d_{int}$).
Comparing variants $T_1$ and $T_2$ we notice the big discrepancy between their interfacial energies (-0.6~J/m$^2$).
The favored $T_1$ structure is characterized by larger Au rumpling parameter
(arising as a result of interactions between interfacial Au atoms with atoms from  the next Ge layer) and smaller interlayer distance than those referred to variant $T_2$.

The Ge-Au bond lengths obtained for $T_1$ structure, 2.50~\AA, are in good agreement with the values taken experimentally for amorphous Ge-Au alloys, 2.66~\AA\, \cite{1991_edwards} and for thin film of Au covering Ge(111) surface, 2.5~\AA\, \cite{1995_over}.
In variant $T_2$ the Au-Ge distance between Au atoms lying directly above Ge  and indicating also $d_{int}$ = 2.36~\AA\ is shorter than Ge-Au bond lengths reported previously.
The preferential localization of Au atoms at bridge sites rather than at top sites of Ge surfaces should be taken into account when modeling other interface model.

We conclude discussion of model hetero-structures with results for
the perfect match \textit{fcc}(111)/\textit{hcp}(001) interface, shown in Fig.~\ref{fig.s001}(c).
It changes the $ABC$ \textit{fcc} arrangement of lattice planes into the $ABAB$ pattern, and,
as can be seen in the figure, one cannot uniquely assign two planes to either of these phases.
We symmetrically divided the gray layers between the two phases and calculated the interface energy values.
As could be expected, the value of the Au-\textit{fcc}/Au-\textit{hcp} interface energy is much smaller (by the order of magnitude) than in the other interfaces.
The work of separation includes also the surface energy and is of the same order of magnitude as in other considered interfaces.

\section{Experimentally observed hetero-structures}
\label{sec.result}

In this Section we present our results of \textit{ab initio} calculations for the Au/Ge interfaces observed in the experimentally grown gold nanoislands on the germanium substrate~\cite{2017_jany}. A particularly regular  \textit{fcc} structure can be found on the Ge(001) surface, with (011) gold plane orientation.
When the sample is annealed above the eutectic temperature, a \textit{hcp} gold phase facing the Ge(111) plane is also observed.

\subsection{Au-\textit{fcc}(011)/Ge(001)}

\begin{figure}[t!]
\centering
\includegraphics[width=\linewidth]{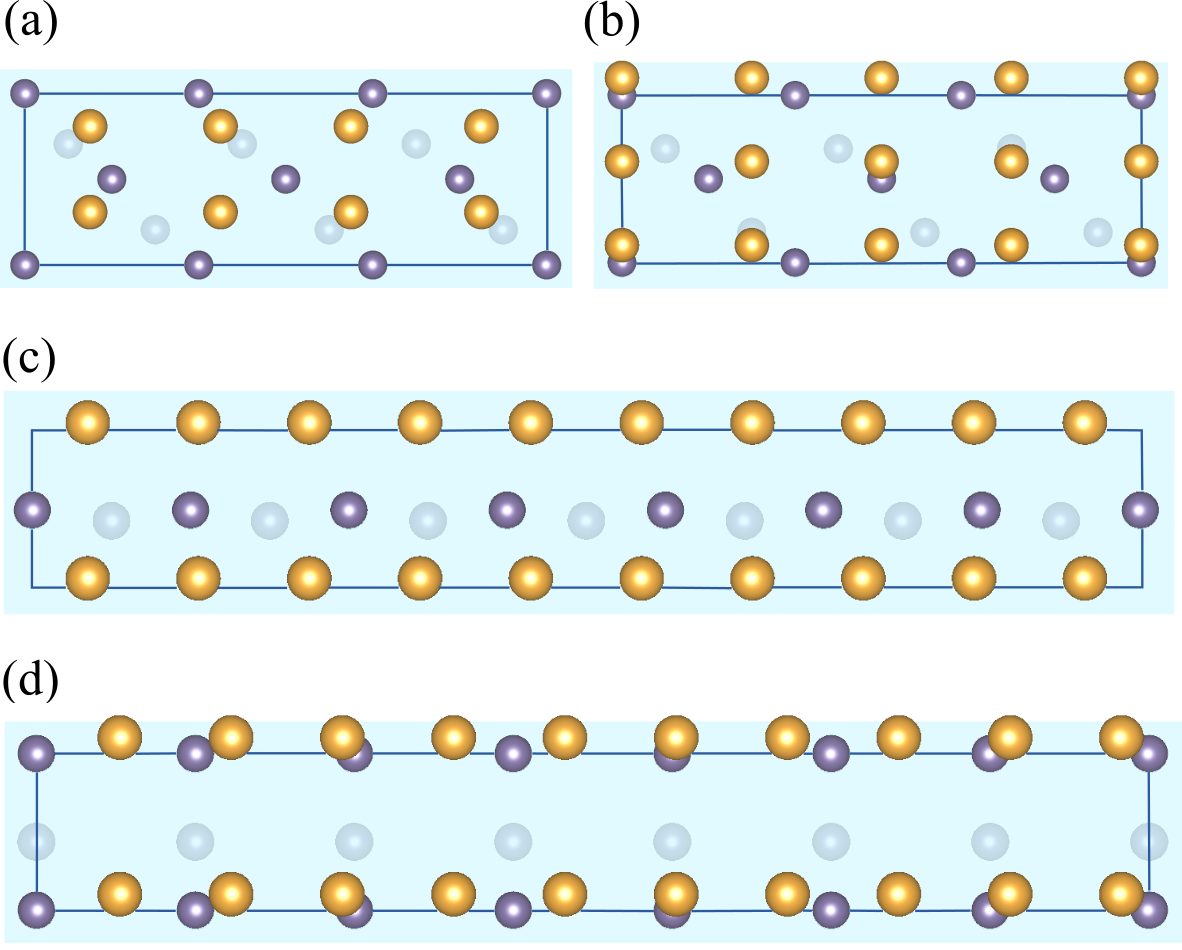}
\caption{Variants $A$, $B$, $C$, and $D$ of the Au-\textit{fcc}(011)/Ge(001) interface (initial configurations) are shown in (a-d) panels, respectively. (a, b) Four Au slabs as shown in Fig.~\ref{fig.basic}(c) are placed upon a Ge supercell constructed from three blocks marked with a black square in Fig.~\ref{fig.basic}(b). In variant $A$ the interface Au atoms avoid Ge atoms, while in  variant $B$ some of the Au atoms are placed directly above Ge atoms. (c, d) Five Au slabs as shown in Fig.~\ref{fig.basic}(c) are joined with seven Ge blocks marked by a green square in Fig.~\ref{fig.basic}(b). Variants $C$ and $D$ of this interface differ in orientation of the Ge slabs with respect to the Au slabs.}
\label{fig.interfaces}
\end{figure}

As already discussed in Sec.~\ref{sec.method}, there are several possibilities of building a low-energy Au-\textit{fcc}(011)/Ge(001) hetero-structure, and here we compare their calculated interface energy values and structural parameters.
First, we join the Au-\textit{fcc}(011) plane oriented as in Fig.~\ref{fig.basic}(a) and the Ge(001) plane from panel \ref{fig.basic}(b), using a cell marked with the black square as a building block for the supercell. In such a way we can get almost strain-free hetero-structure in the $[1\bar10]$Au direction.
In order to reduce strain along the perpendicular [100]Au direction, we take four Au lattice constants and three Ge blocks (the resulting mismatch is about 4\%).
To get benchmark results for this orientation of crystal planes, we consider two different mutual positions of the slabs presented in Figs.~\ref{fig.interfaces}(a) and \ref{fig.interfaces}(b).

In variant $A$, rows of gold atoms are placed between the rows of germanium atoms, and we can expect from previous calculations that
such arrangement is energetically preferable to interfaces in which some Au and Ge atoms are forced to make shorter bonds.
We include five Au and five Ge layers in the supercell, and this gives altogether 70 atoms (30 Ge and 40 Au).
Similarly to the Au(001)/Ge(001) interface, the gold atoms close to the interface  move slightly from their initial positions (rumpling parameter $r_\text{Au}=0.53$~\AA), while the stiffer germanium lattice stays almost unchanged (rumpling parameter $r_\text{Ge}=0.08$~\AA).
In variant $B$, the shift of Au plane  with respect to Ge surface that places the column of Au atoms right over the Ge atoms [Fig.~\ref{fig.interfaces}(b)]
leads to a significant increase in the total energy of the system (supercell) by about 3.5 eV.
Additionally, the forces induced to preserve the Ge-Au bond length on the appropriate level lead to an increase of the rumpling parameter of the Ge layer up to 0.49~\AA.
The interface energy of variant $B$ is higher than of variant $A$ by 0.23~J/m$^2$.
These findings confirm that structures with some Au atoms located directly above the Ge atoms tend to be energetically expensive.

Atomically resolved STEM-HAADF images of gold nanoislands on germanium substrate~\cite{2017_jany} suggest yet another way of matching Au-\textit{fcc}(011) and Ge(001) crystal planes.
Since the distance between germanium atoms in [110] direction closely matches the gold lattice constant (the mismatch of about 1.9\%), we can use a green square in Fig.~\ref{fig.basic}(b) as a new building block for the hetero-structure.

We note that there are two different orientations of this block with respect to the Au-\textit{fcc}(011) slab shown in Fig.~\ref{fig.basic}(c).
Both arrangements are consistent with electron microscopy picture as long as the Au and Ge atoms at the interface are not discriminated.
In the perpendicular direction (with no experimental image to guide us), we chose to join seven germanium blocks and five gold slabs as shown in Fig.~\ref{fig.basic}(c), obtaining a mismatch of about 3\%.
Variant $C$ is constructed in such a way that rows of gold atoms are located between the germanium rows (see Fig.~\ref{fig.interfaces}(c)), and we expect this arrangement to be energetically favored.
Indeed, the interface energy obtained for a supercell containing six layers of Ge (42 atoms) and five layers of Au (50 atoms) is equal to 0.336~J/m$^2$, a value very close to the lowest energy obtained earlier for the optimal Au-\textit{fcc}(001)/Ge(001) structure.

Small displacements of atoms from initial positions are observed in the interface layer ($r_\text{Au}= 0.62$ and $r_\text{Ge}= 0.22$~\AA),
and the distance between Ge and Au interfacial planes is equal to 1.78~\AA\ (again, very similar to our first low energy hetero-structure).
These numbers suggest that variant $C$ is a good candidate for the observed nanostructure.
However, a detailed analysis of the scattered intensity of atomic columns~\cite{2017_jany} suggests that  the rows of Au atoms are located directly above the rows of Ge atoms.
\begin{figure}
\centering
\includegraphics[width=\linewidth]{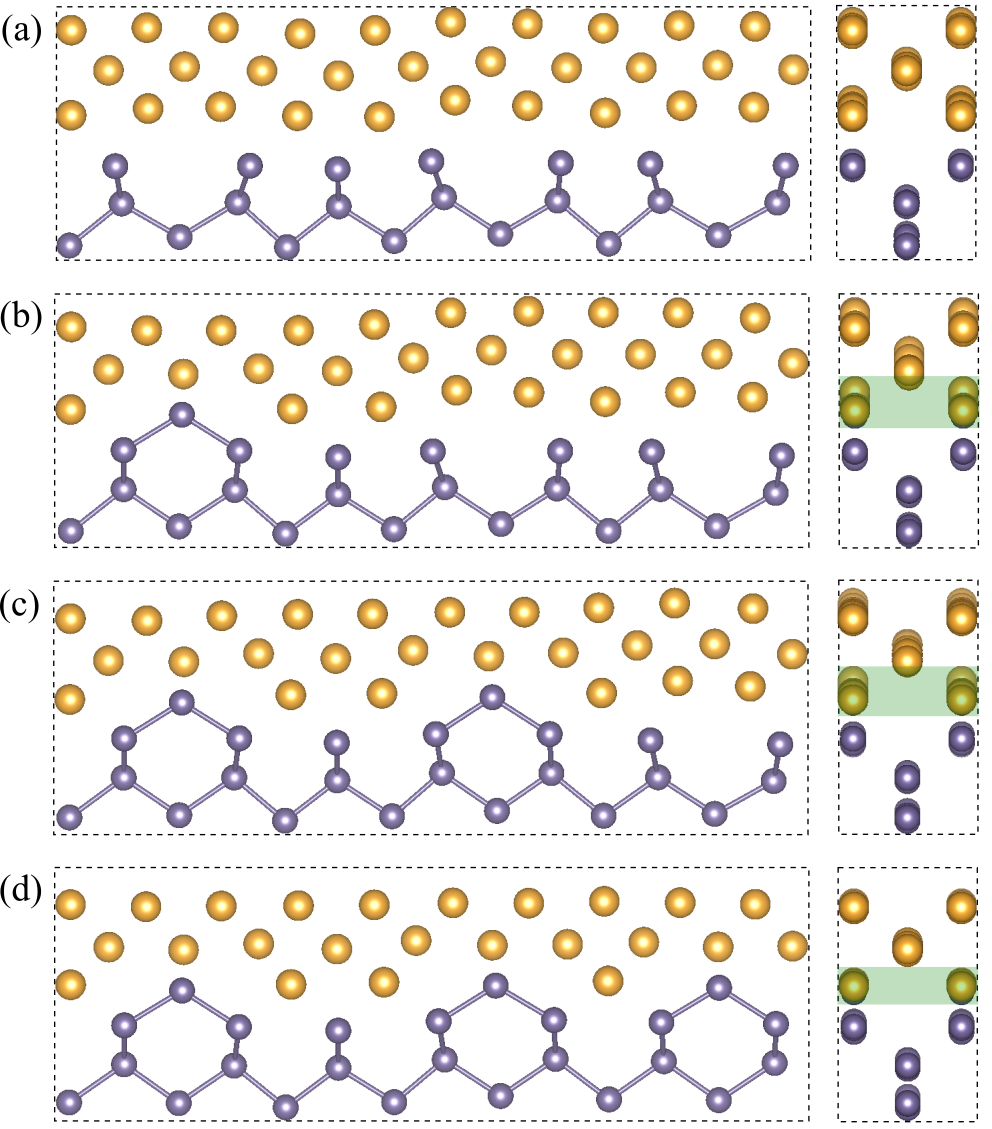}
\caption{
Infusion of defects to variant $D$ of the Au-\textit{fcc}(011)/Ge(001) hetero-structure: (a) the optimized structure of variant $D$; (b-d) additional Ge atoms per supercell introduced to the gold layer from which some Au atoms are removed (twice as many as the number of Ge atoms added). The green rectangle marks the extension of the mixed Au/Ge layer. For better visibility of the changes in the structure, Ge-Ge bonds are shown in the left panels.
}
\label{fig.defects}
\end{figure}
We therefore built variant $D$ of the hetero-structure, with 102 atoms (42 Ge and 60 Au), and interface shown in Fig.~\ref{fig.interfaces}(d).
In order to match the experimental picture, we need a different orientation of the germanium building blocks (rotation by 90$^\circ$ with respect to variant $C$).
Due to incommensurate Au-Au and Ge-Ge distances, it is impossible to find a regular pattern of Au-Ge bonds in the resulting interface layers (some of them might be much shorter than in the optimal Au/Ge arrangement).
After optimization, we get similar values of rumpling parameters as in variant $C$, but the distance between Au and Ge layers is increased by almost 0.4~\AA.
Because of the different numbers of atoms in each supercell we cannot make a direct comparison of the total energies, however, the interface energy in  variant $D$ is higher by approximately 0.1~J/m$^2$.
It is still the second lowest value among variants of the Au-\textit{fcc}(011)/Ge(001)  interface and the difference with respect to the optimized variant $C$ is relatively small.
The slightly higher interface energy in variant $D$ might arise from the distortions
present in the Ge substrate visible in Fig.~\ref{fig.defects}(a), and the enlarged distance between the two phases.

\begin{figure}[t!]
\centering
\includegraphics[width=\linewidth]{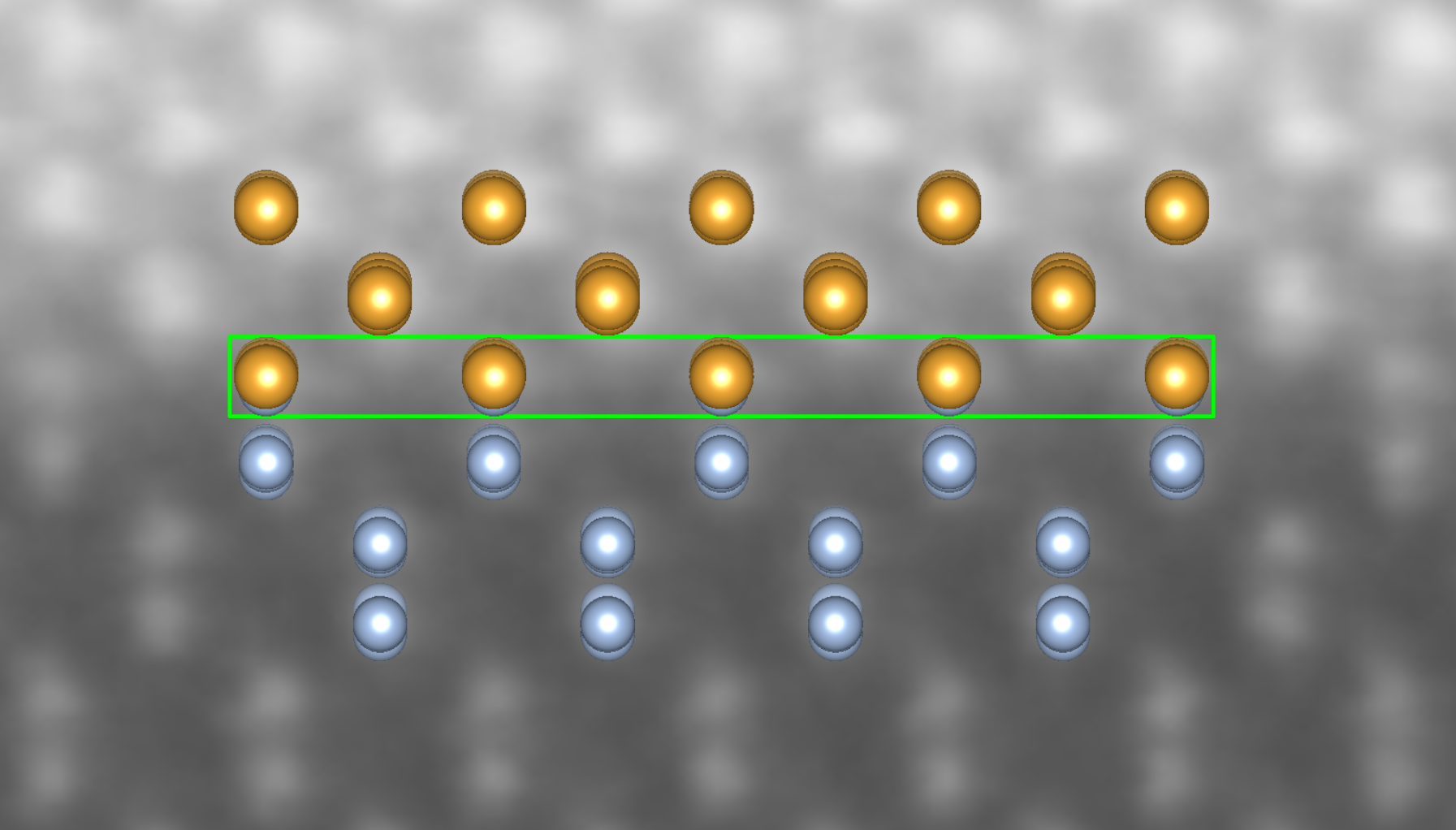}
\caption{
Experimentally observed Au-\textit{fcc}(011)/Ge(001) interface: atomically resolved STEM-HAADF image~\cite{2017_jany}, and  atomic structure obtained for variant $D$ with defects, view as in the right panel of Fig.~\ref{fig.defects}(d), duplicated four times.
The green rectangle marks the mixed layer.
}
\label{fig:fcc_image}
\end{figure}

In the last part of this Section, we discuss a  possibility of lowering the interfacial energy of this variant by introducing defects into the interface layer. %
We investigated three supercells of variant $D$, in which some pairs of Au atoms in the interface layer were replaced by Ge atoms, and their optimized atomic arrangements are presented in Fig.~\ref{fig.defects}(b-d).
We can observe that while one such defect, shown in panel (b), seems to disrupt the arrangement of some atoms even further, more defects restore a remarkably regular structure with a decreased distance between layers at the interface.
The calculated interface energy values for structures in panels (a) to (d) are: 0.437, 0.416, 0.388, and 0.343~J/m$^2$, respectively.
The Ge atoms replacing some Au pairs gradually improve the interface energy and the result obtained for the last modification is very close to the interface energy of variant $C$.
Figure~\ref{fig:fcc_image} shows the experimental image of the investigated interface together with one of our optimized structures, shown in Fig.~\ref{fig.defects}(d).
We note, however, that variant $C$ as well as variant $D$ without defects  would also be consistent with the available experimental image (showing only the view as in right panels of Fig.~\ref{fig.defects}), so we  cannot use it to discriminate these arrangements.

\subsection{Au-\textit{hcp}(010)/Ge(111)}

\begin{figure}[t!]
\centering
\includegraphics[width=\linewidth]{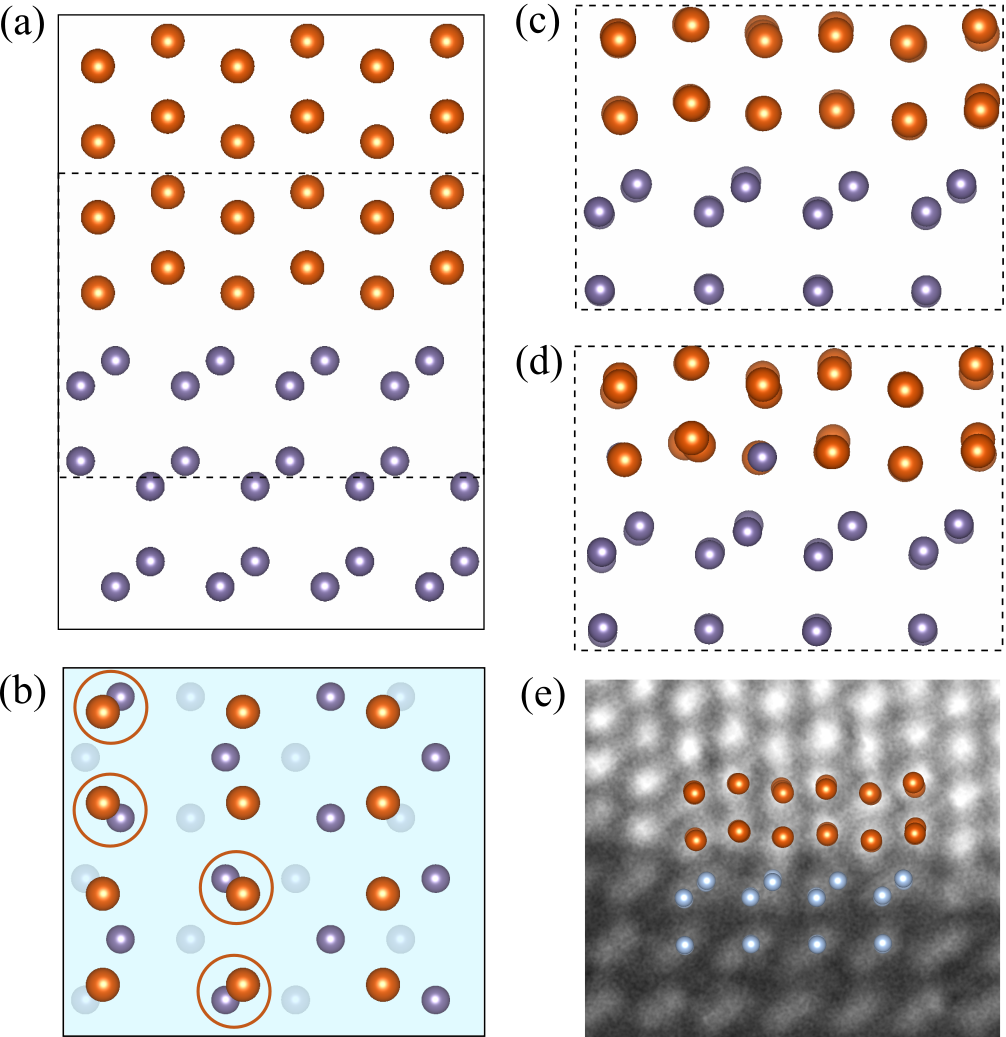}
\caption{The initial configuration of Au-hcp(010)/Ge(111) supercell: (a) side view, and (b) top view of the interface.
The relative position of Au and Ge slabs is selected so as to obtain the maximal possible distance between the nearest Au and Ge atoms.  There are four such short distances per interface (i.e., eight in the supercell) indicated by circles.
(c) The relaxed atomic pattern within the square in panel (a).  (d) The relaxed structure in which Au atoms with the shortest Au-Ge distance are replaced by Ge atoms.
(e) The atomically resolved STEM-HAADF image of the Au-hcp/Ge(111) interface [23] with the optimized supercell as shown in panel (c).}
\label{fig.hcp_model}
\end{figure}

Our last hetero-structure, including the \textit{hcp} phase of gold, is the most interesting one from the perspective of investigating  new phases in nano-structures and their possible applications.
It is however also the most challenging interface for our \textit{ab initio} calculations.
The mutual position of the slabs can be partially read from electron microscopy images~~\cite{2017_jany},
and the identified gold lattice plane adjacent to the Ge(111) surface is (010), with [10$\bar{1}$]Ge parallel to [100]Au-\textit{hcp}.
A relatively low-strain interface can be formed with $4\times3$ Au-\textit{hcp}(010) blocks connected to $3\times2$ Ge(111) blocks as shown in~Fig.~\ref{fig.basic} (the supercell contains 96 Au and 72 Ge atoms).
The Au lattice is extended in the [100]Au-\textit{hcp} direction (with a mismatch of about 3\%) and compressed in the [001]Au-\textit{hcp} direction (a mismatch of 6\%).

In this particular Au/Ge hetero-junction it is not possible to find a starting configuration with similar Au-Ge
distances  across the interface, and inevitably some of them are significantly shorter if compared to other optimized structures.
In the interface used for optimization, presented in Fig. \ref{fig.hcp_model}(a, b), the mutual position of the slabs is chosen in such a way that the shortest \mbox{Au-Ge}
 bonds are maximal for a given distance between slabs, i.e. extremely short bonds are avoided.
It occurs that at each interface there are four short Au-Ge bonds, marked with circles in Fig.~\ref{fig.hcp_model}(b).
As a consequence, DFT calculations with full atomic displacements lead to strong changes in the Au atomic positions.
In the he optimized structure shown in Fig.~\ref{fig.hcp_model}(c) the characteristic $hcp$ zigzags are flattened.
The lattice strain present in the Au part of the hetero-structure may also be a factor contributing to atomic displacements away from the \textit{hcp} lattice positions.

The obtained value of interface energy (0.369~J/m$^2$) is similar to the previously investigated variants $C$ and $D$ of Au-\textit{fcc}(011)/Ge(001) hetero-structure.
The work of separation (1.506~J/m$^2$)
is smaller due to different surface energies of the respective building blocks.
\begin{figure}[t!]
\centering
\includegraphics[width=1.04\linewidth]{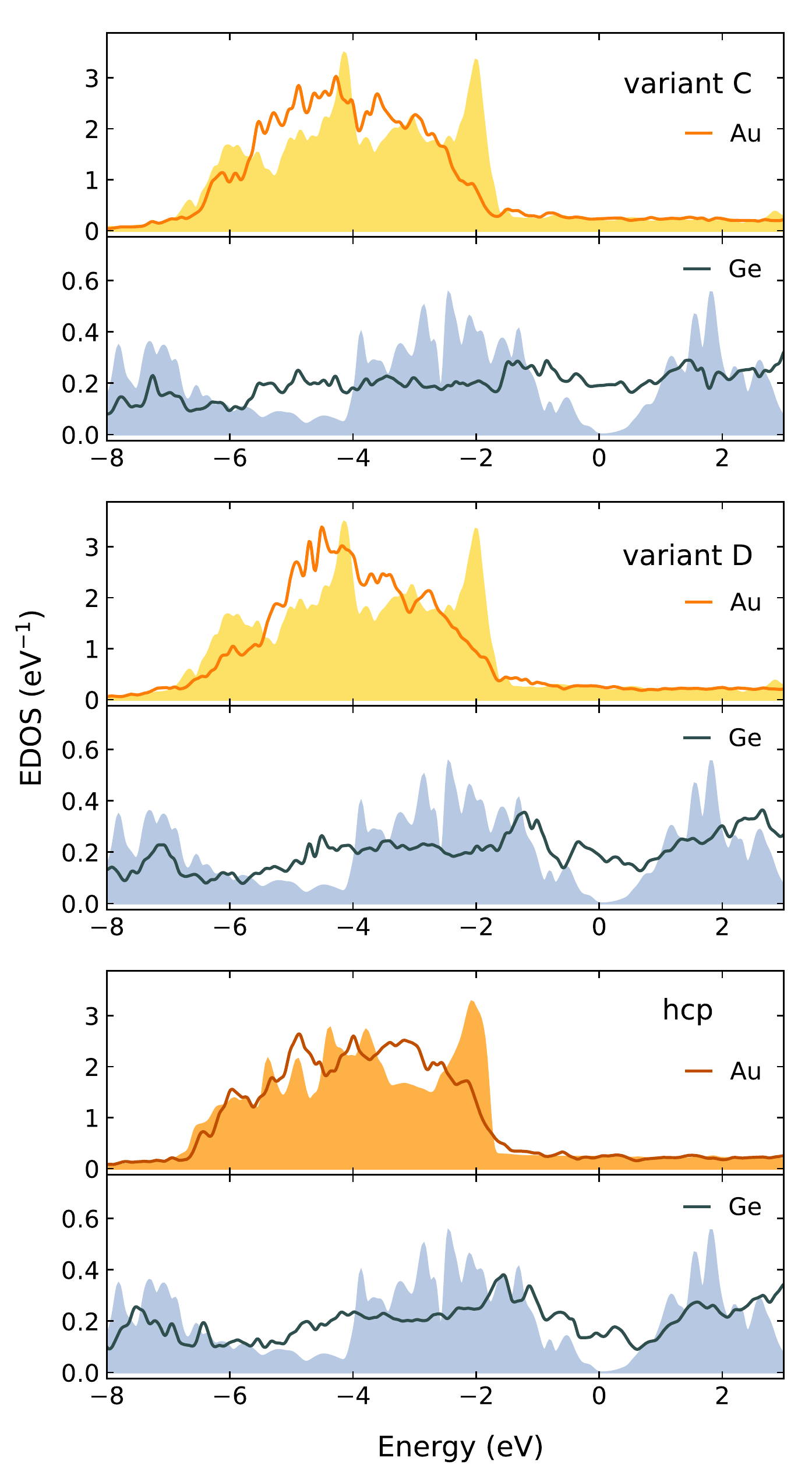}
\caption{
The electron density of states (EDOS) for Au and Ge atoms, averaged over all atoms in layers adjacent to the interface. The results for the respective bulk phases are presented as filled areas. For all plots $E_F = 0$.
}
\label{fig.edos}
\end{figure}

We considered also possible defects in the lattice that could lower the energy cost of the short \mbox{Au-Ge} bonds and better preserve the \textit{hcp} ordering in the Au part of the slab.
In the structure shown in Fig.~\ref{fig.hcp_model}(b), we introduced Ge atoms in place of Au atoms in the four positions at the interface where the Au-Ge distances are minimal (marked with circles).
While the \textit{hcp} zigzags are more pronounced in the optimized supercell shown in panel (d), the interface energy is slightly higher (0.395~J/m$^2$) with respect to defect-free junction.
We also tested another type of defects (Au vacancies). To construct the supercell, we first found the position of slabs with the smallest possible Au-Ge distances (two such pairs per interface can be found), and removed the Au atoms from these bonds.
The optimized lattice preserves the hexagonal arrangement, however, the interface energy is significantly higher than in other considered \textit{hcp} variants (0.514~J/m$^2$).

Figure~\ref{fig.hcp_model}(e) shows the experimental picture and the defects-free optimized structure shown in panel (a).
We can observe that the atomic pattern at the interface is well reproduced by our model.
The interface seen in the experimental image is not as regular as the Au-\textit{fcc}/Ge junction and other types of defects may additionally stabilize this heterostructure.
%
\begin{figure}[b!]
\centering
\includegraphics[width=0.9\linewidth]{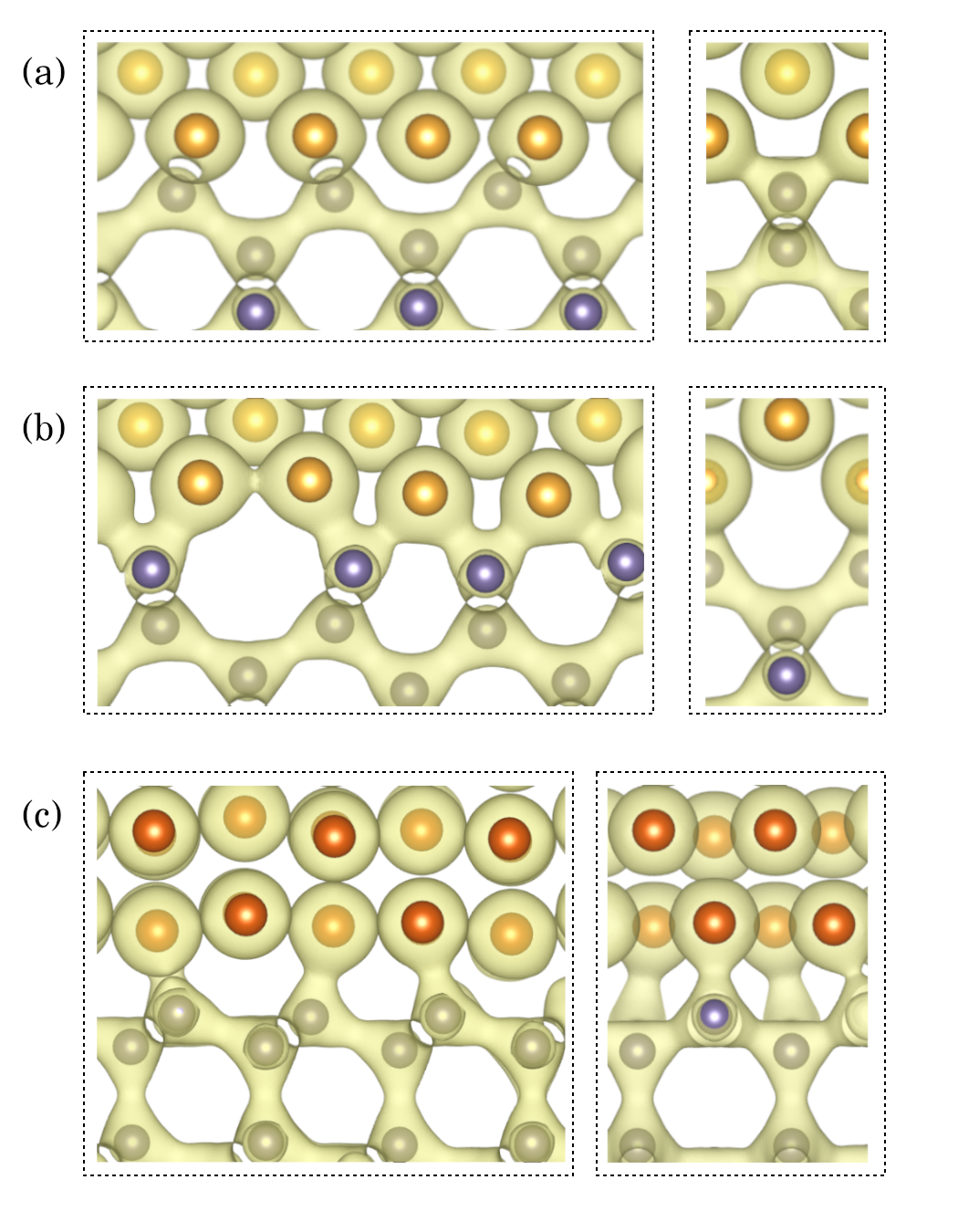}
\caption{
The isosurfaces of charge density (0.051~e/\AA$^{-3}$) for variants: (a) $C$ and (b)~$D$~of~Au-\textit{fcc}(011)/Ge(001) interface, compared with the isosurface of Au-\textit{hcp}(010)/Ge(111) interface (c).
}
\label{fig.chg}
\end{figure}

\section{Electronic structure}
\label{sec.electronic}

Here we discuss electronic properties for optimized structures of the defect-free Au/Ge interfaces investigated in the previous section, i.e.
variants $C$ and $D$ of the Au-\textit{fcc}(011)/Ge(001) interface and the Au-\textit{hcp}(010)/Ge(111) heterostructure shown in Fig.~\ref{fig.hcp_model}(a).
Figure~\ref{fig.edos} presents the electron density of states (EDOS) for Au and Ge atoms (averaged over all atoms in layers adjacent to the interface).
These results are compared to the bulk density of states (we note that small differences in the bulk phase between Au-\textit{fcc} and Au-\textit{hcp} can be seen from the plots).
First of all, electron density at the Fermi level is finite for the Ge atoms, making the interface region metallic (we observe this effect also for other Ge layers with only slight decrease in the density of states with the distance from the interface).
On the other hand, the EDOS for gold atoms remains essentially unchanged around the Fermi level with respect to the bulk phase for all investigated interfaces.
However, the big peak near -2~eV disappears and the density is shifted towards lower energy values. %
This shift can be understood as formation of bonding states with germanium electrons.

\begin{figure}[t!]
{\centering
\includegraphics[width=1.\linewidth]{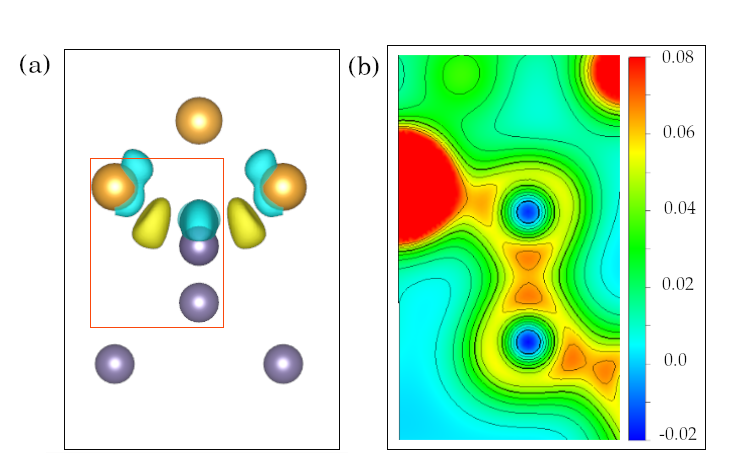}}
\caption{
(a) The isosurfaces of the charge density difference for variant $C$ at 0.0064~e/\AA$^{-3}$ (yellow and blue correspond to positive and negative values, respectively). (b) The map of charge density distribution for a plane defined by three atoms in a red frame of panel (a). The orientation of this plane is different from the view in panel (a).
}
\label{fig.diff-map}
\end{figure}

Emergence of Ge-Au bonding in the modeled hetero-structures can be illustrated by plots of charge distribution at the Au/Ge interface.
Fig.~\ref{fig.chg} presents isocharge surfaces (0.051 e/\AA$^{-3}$),
connecting pairs of atoms if a bond is formed between them.
Away from the interface, each germanium atom shares four covalent bonds with its nearest neighbors, and gold atoms form a metallic phase.
At the interfaces, bonding states can also be seen between Ge and Au atoms:
for Au-\textit{fcc}(011)/Ge(001) interfaces each Ge atom is attached to two Au atoms, while the Ge atoms at the Au-\textit{hcp}(010)/Ge(111) interface are bonded with one Au atom
(the difference arises from different number of unbound germanium electrons in a given crystal surface).

Even deeper insight into formation of these bonding states can be gained by investigating the charge density and charge redistribution at the interface.
Figure~\ref{fig.diff-map}(a) shows the charge accumulation (yellow) and  charge depletion (light blue) regions with respect to separate Ge and Au slabs terminated with vacuum (see Appendix B for details).
One can observe charge accumulation between the interface Ge and Au atoms, as expected for bond formation.
Panel (b) of Fig.~\ref{fig.diff-map} presents the charge density map for a plane defined by three atoms inside a red frame in panel~(a), i.e. a cross-section through both Au-Ge and Ge-Ge bonds, revealing their detailed structure.
Perhaps unexpectedly, the charge density between the interface Au and Ge atoms reaches almost the same level as for Ge-Ge covalent bonds, see Fig.~\ref{fig.diff-map}(b).
The integrated charge density of the bonding electrons, however, can be expected to be significantly smaller, one can also observe that  Au-Ge bond is strongly asymmetrical. %
Similar bonding states were already observed for Au atoms deposited on Ge(001) surface using the X-ray photoelectron spectroscopy spectra, supported with first-principles calculations
\cite{popescu2013}.

\section{Summary and conclusions}
\label{sec.summa}

In summary, we have used density functional theory to study the structure and stability of phase boundaries found in gold nanoislands grown on the Ge substrate.
The atomically resolved experimental images report the sharp Au-\textit{fcc}/Ge junction of near perfect Au-\textit{fcc} and Ge structures.
The observed Au-\textit{hcp}/Ge interface is uneven, with several steps in the Ge(111) substrate a few nanometers apart. However, fragments of a regular junction between constituent lattices are still visible between the steps and thus we considered interface models without dislocations. However, we discussed possibilities of point defects and vacancies at the interface layer.

We focused on the interfacial energy values that in our approach mostly capture the bonding energy of the interface atoms.
The calculations for the Au-\textit{fcc}(011)/Ge(001) interface reveal the lowest interfacial energies for variants $C$ and $D$ of hetero-junctions consistent with available microscopic images.
Refined experimental data, indicating that variant $D$ (with slightly higher interface energy) is realized, led us to search for possible defects in the interface layer that might lower the energy cost of this junction.
The optimized supercell is characterized by both a low  interfacial energy and a regular atomic pattern fitting the experimental picture.
The presented approach provides helpful insight into structure and energy costs of deformations present in several investigated variants.

We also constructed several supercells for studying the novel \textit{hcp} phase of gold.
Optimization of the atomic positions in the superlattice shows significant atomic displacements that can result mainly from uneven distribution of Au-Ge distances.
Again, the creation of defects might stabilize the observed Au-\textit{hcp}(010)/Ge(111) structure by eliminating part of the internal stress.

Electronic structure suggests that the interface region becomes metallic but bonding states between the interface atoms are also formed. The detailed electron density maps show that the charge transfer is present both at the interface and in the adjacent germanium layer.

\begin{acknowledgments}
The authors gratefully thank Johan Verbeeck and Nicolas Gauquelin
from EMAT University of Antwerp for atomically resolved HAADF
STEM imaging of nanoislands made for $fcc$ and $hcp$ phases of gold.
The authors kindly acknowledge support by the National Science Centre
(NCN, Poland) under Projects No.: 2017/25/B/ST3/02586 (P.P. and M.S.),
2020/04/X/ST5/00539 (B.R.J.), and 2021/43/B/ST3/02166 (A.M.O.). A.M.O.
is grateful for support via the Alexander von Humboldt \mbox{Foundation}
\mbox{Fellowship} \cite{AvH} \mbox{(Humboldt-Forschungspreis).} This research was supported in part by the Excellence Initiative -- Research University Program at the Jagiellonian University in Krak\'ow.

Some figures in this work were rendered using {\sc Vesta} software
\cite{vesta}.
\end{acknowledgments}

\appendix
\section{Surface energies}
\label{sec.appA}

The surface energies calculated for several planes of Ge and Au-\textit{fcc} crystals are given and compared to available theoretical and experimental data in Table~\ref{tab.surfener}.
We also present results for the \textit{hcp} phase of gold, considering the hexagonal $(001)$ plane as well as the (010) plane relevant for our experimental interface models.
It is worth noticing that the differences between Au-\textit{fcc}(111) surface and  Au-\textit{hcp}(001) surface begin from the third layer only and this explains why the numerical results are almost identical for these planes.
Generally, a surface with lower energy is usually considered more stable since it needs smaller amount of work to be done when cleaving along this surface.
Thus, the Ge(111) and Au-\textit{fcc}(111) free surfaces are more favored than others.
\begin{table}[h!]
\caption{Surface energies calculated for various low-index planes of Ge, Au-\textit{fcc} and Au-\textit{hcp} crystals. For comparison previous theoretical results  and experimentally derived values are presented. }
\label{tab.surfener}
\begin{ruledtabular}
\begin{tabular}{l c c c}
 & \multicolumn{3}{c}{surface energy (J/m$^2$) } \\
\colrule
Ge-diamond  &~~ (001) ~~ &~~ (011) ~~& ~~(111)~~ \\
this work                    & 1.394    & 1.241     &  1.035     \\
theory (LDA) \cite{Stekolnikov2002}  & 1.710    & 1.510 & 1.320 \\
exp \cite{Jaccodine1963}       & 1.835    & 1.300 & 1.060 \\
\hline
Au-\textit{fcc} & ~~ (001) ~~ &~~ (011) ~~& ~~(111)~~ \\
this work & 0.813    &    0.834    & 0.641      \\
theory (GGA) \cite{Holec2020}  & 0.87 & 0.91 & 0.72 \\
theory (LDA) \cite{Holec2020}  & 1.34 & 1.43 & 1.26 \\
exp \cite{Tyson1977}    &-- &-- & 1.5\\
\hline
Au-\textit{hcp} & (001) &  (010)  & \\
this work       & 0.639 &  0.840  & \\
\end{tabular}
\end{ruledtabular}
\end{table}

\begin{figure}[]
\centering
\includegraphics[width=0.9\linewidth]{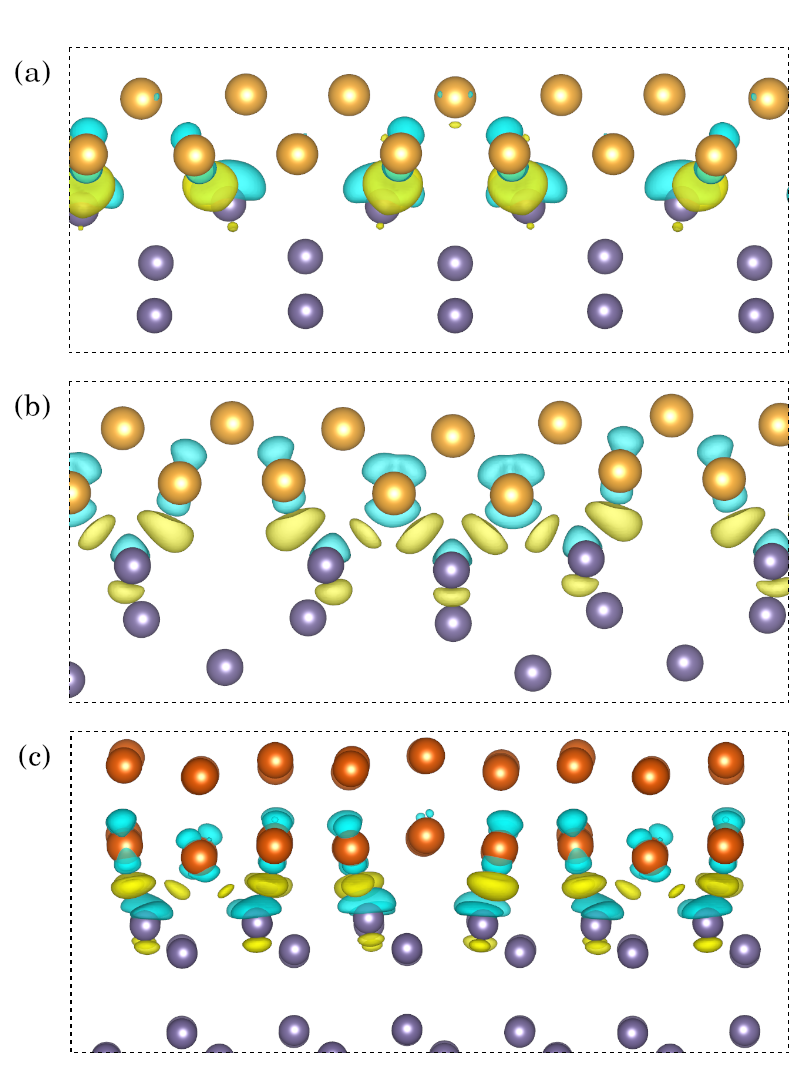}
\caption{
The charge density differences (0.0045~e/\AA$^{-3}$) for variants $C$
(a) and $D$ (b) of Au-\textit{fcc}(011)/Ge(001) interface compared with the isosurface of Au-\textit{hcp}(010)/Ge(111) interface (c). Yellow and cyan regions indicate the electron
accumulation and depletion, respectively.
}
\label{fig.diff}
\end{figure}

\section{Charge density differences}
\label{sec.appB}

The changes in charge distribution induced by the Au/Ge interface can be demonstrated by calculating the difference between the total charge density of the interface ($\rho_{\mathrm{Au/Ge}}$) and charge densities of systems with vacuum substituting for Au ($\rho_{\mathrm{vac/Ge}}$) or Ge ($\rho_{\mathrm{Au/vac}}$) in accordance with the formula:
\begin{eqnarray}
\Delta \rho = \rho_{\mathrm{Au/Ge}} - \rho_{\mathrm{vac/Ge}} - \rho_{\text{Au/vac}}.
\label{eq:rho}
\end{eqnarray}
The charge density differences calculated for variant $C$ and $D$ of the  Au-\textit{fcc}(011)/Ge(001) interface and for the  Au-\textit{hcp}(010)/Ge(111) interface are illustrated in Fig.~\ref{fig.diff} with yellow regions representing charge accumulation and light blue regions indicating charge depletion.
For all heterostructures, the charge transfer from  Au and Ge intefacial atoms leads to the electron accumulation at approximately half of their distance.
At the presented isocharge surface value, the enhancement of charge density between interfacial Ge atoms and their nearest neighbors in the substrate is also observed for variant D and the Au-\textit{hcp}/Ge  interface.

\newpage
%

\end{document}